\newcommand*{\ATLASLATEXPATH}{}
\documentclass[cernpreprint, texlive=2016, USenglish]{\ATLASLATEXPATH atlasdoc}
 
\usepackage{\ATLASLATEXPATH atlaspackage}
\usepackage{\ATLASLATEXPATH atlasbiblatex}
 
\usepackage{\ATLASLATEXPATH atlasphysics}
 
\usepackage{float}
\usepackage{multirow}
 
\graphicspath{{logos/}{figures/}}
 
\usepackage{rotating}
\usepackage{amssymb}
 
\usepackage{ANA-HIGG-2018-54-PAPER-defs}

\AtlasTitle{Combination of searches for invisible Higgs boson decays with the ATLAS experiment}
 
\AtlasAbstract{
Dark matter particles, if sufficiently light, may be produced in decays of the Higgs boson.
This Letter presents a statistical combination of searches for $H\to\text{invisible}$ decays where $H$ is produced according to the Standard Model via vector boson fusion, $Z(\ell\ell)H$, and $W\!/\!Z(\text{had})H$, all performed with the ATLAS detector using 36.1~fb$^{-1}$ of $pp$ collisions at a center-of-mass energy of $\sqrt s = 13$~TeV at the LHC.
In combination with the results at $\sqrt s = 7$ and 8~TeV, an exclusion limit on the $H\to\text{invisible}$ branching ratio of  $0.26~(0.17^{+0.07}_{-0.05})$ at 95\% confidence level is observed~(expected).
}
 
\author{The ATLAS Collaboration}
 
\AtlasRefCode{HIGG-2018-54}

\AtlasCoverSupportingNote{Paper draft}{https://cds.cern.ch/record/2644319}
\AtlasCoverSupportingNote{Support note}{https://cds.cern.ch/record/2320415}

\AtlasCoverAnalysisTeam{Oleg Brandt, Song-Ming Wang, Hideki Okawa, Patrick Rieck, Yusheng Wu, Jike Wang, Iouri Naryshkin, Anna Shcherbakova, Loan Truong, Wei Wang, Rui Zou, Cong Geng, Paul Philipp Gadow, Benjamin Taylor Carlson, Sascha Simon Dreyer, Makoto Teshima}
 
\AtlasCoverEdBoardMember{Pierre-Hugues, Beauchemin (Tufts) (chair)}
\AtlasCoverEdBoardMember{Lydia, Iconomidou-Fayard (Orsay LAL) }
\AtlasCoverEdBoardMember{Paolo, Francavilla (Pisa) }

\AtlasCoverEgroupEditors{atlas-HIGG-2018-54-editors@cern.ch}
 
\AtlasCoverEgroupEdBoard{atlas-HIGG-2018-54-editorial-board@cern.ch}
 
\PreprintIdNumber{CERN-EP-2019-046}
\AtlasJournal{Phys.\ Rev.\ Lett.}
 
\AtlasJournalRef{Phys. Rev. Lett. 122 (2019) 231801}
 
\AtlasDOI{10.1103/PhysRevLett.122.231801}
\hypersetup{pdftitle={ATLAS document},pdfauthor={The ATLAS Collaboration}}
 
\begin{document}
 
\maketitle

One of the central open questions in physics today is the
nature of dark matter (DM) that is found to comprise most of the matter in
the universe~\cite{Akrami:2018vks,Hinshaw:2012aka,Bertone:2004pz,Trible:dmu}. A compelling candidate for DM is a
stable electrically neutral particle $\chi$ whose non-gravitational
interactions with Standard Model (SM) particles are weak.
Such a particle with a mass comparable to the mass scale of the electroweak sector particles could be detectable~\cite{Goldberg:1983nd,Ellis:1983ew,Jungman:1995df}
and accommodate the observed DM relic
density~\cite{Steigman:1984ac,Scherrer:1985zt}. Numerous models predict
detectable production rates of such DM particles at the Large Hadron
Collider (LHC)~\cite{Abercrombie:2015wmb,Feng:2010gw,Boveia:2018yeb}. In a wide class of those models,
the 125~GeV Higgs boson $H${\mbox{~\cite{Aad:2012tfa,Chatrchyan:2012ufa}} acts as a portal between a dark sector and the SM sector, either through Yukawa-type couplings to fermionic dark matter, or other mechanisms~\cite{Antoniadis:2004se,ArkaniHamed:1998vp,Datta:2004jg,Kanemura:2010sh,Djouadi:2011aa,Djouadi:2012zc,Shrock:1982kd,Choudhury:1993hv,Eboli:2000ze,Davoudiasl:2004aj,Godbole:2003it,Ghosh:2012ep,Belanger:2013kya,Curtin:2013fra}. If kinematically allowed, decays of the Higgs boson to DM particles represent a distinct signature in such models. Higgs boson decays to DM particles can only be indirectly inferred through
missing transverse momentum\footnote{ATLAS uses a right-handed coordinate
system with its origin at the nominal interaction point (IP) in the
center of the detector and the $z$-axis along the beam pipe. The
$x$-axis points to the center of the LHC ring, and the
$y$-axis points upward. Cylindrical coordinates $(r,\phi)$ are used
in the transverse plane, $\phi$ being the azimuthal angle around the
$z$-axis. The pseudorapidity is defined in terms of the polar angle
$\theta$ as $\eta=-\ln\tan(\theta/2)$. The distance between two
objects in $\eta$--$\phi$ space is $\Delta R = \sqrt{(\Delta \eta)^2
+ (\Delta \phi)^2}$. Transverse momentum is defined by $\pt =
p\sin{\theta}$.}
\met due to DM particles escaping detection, and are therefore termed ``invisible''~(inv).
 
Direct searches for invisible Higgs boson decays have been carried out with the ATLAS detector~\cite{Aad:2008zzm,Capeans:2010jnh,Abbott:2018ikt} in Run~1 of the LHC, using up to 4.7~\ifb of $pp$ collision data at a center-of-mass energy of $\sqrt{s} = 7~\TeV$ and up to 20.3~\ifb\ at 8~TeV. Different event topologies were considered,  assuming SM production rates: vector boson fusion (VBF)~\cite{HIGG-2013-16}, Higgsstrahlung from a $Z$ boson decaying into a pair of electrons or muons (\zh)~\cite{HIGG-2013-03}, and Higgsstrahlung from a $W$ or $Z$ boson decaying into hadrons (\vh)~\cite{HIGG-2014-07}.
These searches for invisible Higgs boson decays have been statistically combined, and an upper limit at 95\% confidence level~(CL) on the invisible Higgs boson branching ratio of $\brinv~<~0.25~(0.27^{+0.10}_{-0.08})$~\cite{bib:run1paper} was observed (expected). In combination with visible decay modes of the Higgs boson, the upper observed (expected) limit improved to 0.23~(0.24)~\cite{bib:run1paper}.
Direct searches for invisible Higgs decays were performed using up to 36.1~\ifb\ of $pp$ collision data at $\sqrt{s} = 13~\TeV$ recorded in 2015 and 2016 in the VBF~\cite{bib:vbfpaper}, $\zh$~\cite{bib:zhpaper}, and $\vh$~\cite{bib:vhpaper} topologies at ATLAS.
The aforementioned results at $\sqrt{s}=13~\TeV$ will be referred to as ``Run~2 results'' in the following. Similar searches were performed by the CMS Collaboration~\cite{Sirunyan:2018owy,Sirunyan:2017qfc,Sirunyan:2017jix,Khachatryan:2016whc,Chatrchyan:2014tja}.

This Letter presents the statistical combination of the Run~2 searches with 36.1~\ifb of data for invisible decays of the Higgs boson using the ATLAS detector.
Subsequently, a statistical combination with the combined Run~1 result~\cite{bib:run1paper} from ATLAS is performed. An overview of all results used as inputs in this combination is given in Table~\ref{tab:resultsall}. The analysis is performed under the assumption of SM Higgs boson production. Visible decay modes of the Higgs boson are not considered.

A brief overview of the Run~2 searches for \hinv is given below.
\begin{itemize}
\item
\textbf{VBF topology~\cite{bib:vbfpaper}:\quad}
The analysis of the VBF production mode employs a \met trigger that is 98\% efficient or better in the considered region of phase space.
The event selection requires~$\met~>~180~\GeV$.
Jets ($j$) are reconstructed up to $|\eta(j)| < 4.5$ from energy clusters in the calorimeter
using the anti-$k_{t}$ algorithm~\cite{Cacciari:2008gp} with a radius parameter $R=0.4$.
The two jets leading in \pt are required to be separated by $|\Delta\eta_{jj}| > 4.8$.
There should be no additional jets with $\pt > 25$ GeV and no isolated electron or muon candidate with $\pt > 7$ GeV.
These requirements serve to reduce the contribution from $W/Z$ production in association with jets (\vjets).
In the search signal region (SR) the $\mjj$ distribution of the background falls more rapidly than the signal,
where $\mjj$ represents the invariant mass of the two selected leading jets.
Thus the SR is divided into three $\mjj$ regions ($1<\mjj/\TeV<1.5$, $1.5<\mjj/\TeV<2$, and $\mjj/\TeV>2$)
to improve the search sensitivity.
The dominant background sources are $Z(\nu\nu)+\text{jets}$ and
$W(\ell\nu)+\text{jets}$ production, where the charged lepton $\ell$ is not detected.
Control regions (CR) enriched in $Z(\ell\ell)+\text{jets}$ and
$W(\ell\nu)+\text{jets}$ processes with $\ell=e,\mu$ are defined to determine the respective normalization factors in the SR.
The main contributions to uncertainties are from the finite number of simulated Monte Carlo (MC) events, the modelling of \vjets production, and accuracy of the jet energy scale~(JES).
The final discriminant is the number of events in the three $\mjj$ regions.
\item
\textbf{$\boldsymbol{\zh}$ topology~\cite{bib:zhpaper}:\quad}
This search is conducted in the Higgsstrahlung channel where the $Z$ boson decays into a pair of electrons or muons.
A selected candidate event must pass at least one of the various single-lepton triggers, fulfill $\met > 90$~GeV and $\met/H_T > 0.6$, where $H_T$ is calculated as the scalar sum of the \pt of the selected leptons and jets,
and have exactly one pair of isolated electrons or muons with an invariant mass that is consistent with that of the $Z$ boson.
The transverse momentum requirement on the leading (subleading) charged lepton is $\pt > 30$~(20)~GeV.
To reduce the \zjets background, the dilepton system must be aligned back-to-back relative to the $\met$ vector  in the transverse plane.
Events with jets originating from $b$-quarks ($b$-jets) are vetoed to suppress backgrounds from
top quark pair ($t\bar{t}$) production and $W$ boson production in association with a single top quark ($Wt$).
The irreducible $Z(\nu\nu)Z(\ell\ell)$ background is estimated from MC simulations and its production yield is normalized to the
theoretical prediction of Ref.~\cite{Grazzini:2015hta,Caola:2015psa}.
The $W(\ell\nu)Z(\ell\ell)$ background contribution is also predicted with MC simulations and is normalized by a scale factor
that is obtained from a CR enriched in $WZ$ events.
The \zjets background is estimated with a data-driven method that uses $Z$-enriched CRs.
The final discriminant is \met.
\item
\textbf{$\boldsymbol{\vh}$ topology~\cite{bib:vhpaper}:\quad}
This analysis considers the Higgsstrahlung channel where the associated $W$ or $Z$ boson decays into hadrons.
The final state signature of large $\met$ and jets also receives contributions from
Higgs boson production via gluon fusion with jets originating from
initial state radiation, and production via the VBF process.
Selected events must pass a \met trigger and must not contain an isolated electron or muon with $\pt > 7~\GeV$.
As a $V$ is boosted, the two jets from its decay become increasingly collimated and are eventually merged into one single reconstructed jet.
Thus, this search is conducted in two topological channels.
In the ``merged'' topology, the SR is defined with~$\met~>~250$~GeV and has at least one trimmed~\cite{Krohn2010,ATLAS-CONF-2016-035}
large-$R$ jet~($J$) that is reconstructed using the anti-$k_{t}$ algorithm with $R=1.0$.
The signal large-$R$ jet is the one with the highest \pt.
For the ``resolved'' topology, the selected event should have $\met > 150$ GeV and at least two
small-$R$ jets~($j$) with $R=0.4$.
Each event is first passed through the merged topology selection and, if it fails,
it is passed through the resolved topology selection.
To improve the search sensitivity, the selected events are further split into
categories with zero, one, and two identified $b$-jets, and into two mass regions of the invariant mass of the
signal large-$R$ jet (two signal small-$R$ jets) for the merged (resolved) topology.
The low mass region ($70 \lesssim m_{J}, m_{jj}/\GeV \lesssim 100$) targets the hadronic $W$/$Z$ boson decays
of the associated production, whereas the high mass region ($100 \lesssim m_{J}, m_{jj}/\GeV <250$) is sensitive to  gluon fusion and VBF production.
The main background contributions are from the \vjets and $t\bar{t}$ processes.
The predictions from MC simulations are constrained with CRs
that contained one or two leptons, and are kinematically similar to the SR.
The final discriminant is $\met$.
\end{itemize}
The SRs and CRs of the individual input analyses are either orthogonal by construction,
or were shown to have an overlap below 1\%, which is neglected in the following.
\enlargethispage{1cm}

The statistical combination of the analyses is performed by constructing the product of their likelihoods and maximizing the resulting likelihood ratio $\Lambda(\brinv;\theta)$~\cite{Cowan:2010js}. This is done following the implementation described in Ref.~\cite{Verkerke:2003ir,Moneta:2010pm}, with \brinv as the parameter of interest.  Systematic uncertainties are modeled in the likelihood function as nuisance parameters $\theta$ constrained by Gaussian or log-normal probability density functions~\cite{bib:run1paper}. Expected results are obtained using the Asimov dataset technique~\cite{Cowan:2010js}.

In the combination of Run~2 results, most experimental systematic uncertainties as well as the uncertainty on the integrated luminosity and the modeling of additional $pp$ collisions in the same and neighboring bunch crossings~(pile-up) are correlated across all search channels. Some experimental uncertainties related to flavor tagging and the JES
are represented through different parameterizations in the input analyses and are therefore treated as uncorrelated. The impact of this assumption on the combined result is estimated using alternative correlation models where the leading sources of systematic uncertainty in the respective parameterizations are treated as correlated, and found to have an absolute effect on the \brinv limit of the order of $0.01$. The systematic uncertainties on the total \hinv signal cross section due to the choice of parton distribution functions (PDF) are considered correlated among all channels. By contrast, uncertainties due to missing higher order corrections are estimated through variations of factorization and renormalisation scales and treated as correlated between the \zh and \vh processes. This is not done for VBF, which represents a distinct topology.
The impact of the corresponding uncertainties on the acceptance rather than the total cross section of \vh production is evaluated and found negligible.
Few systematic uncertainties that are tightly constrained in a given analysis are left uncorrelated in order not to introduce any potential phase space specific biases.

The negative logarithmic profile likelihood ratios $-2\,\Delta\!\ln(\Lambda)(\brinv;\theta)$ as a function of \brinv of the individual analyses and of the combined Run 2 result are shown in Fig.~\ref{fig:lhscan}, corresponding to a best-fit combined value of $\brinv=0.20\pm0.10$.
The dominant uncertainty sources are finite event yields in data and MC simulations, reconstruction of jets and leptons, and modelling of diboson and $W/Z+\text{jets}$ production. In absence of a significant excess, an upper limit at 95\% CL of $\brinv < \result$ is observed (expected) with the $CL_s$ formalism~\cite{Read:2002hq} using the profile likelihood ratio as a test statistic. The excess in data corresponds to a $p_\text{SM}$-value of 3\% under the SM hypothesis of $\brinv\simeq10^{-3}$, and is a direct consequence of the excesses that are present in each of the three input analyses, see Table~\ref{tab:resultsall}. Each of the individual analyses has been scrutinized and these excesses have been found non-significant and independent.

\begin{figure}
\centering
\includegraphics[width=0.45\textwidth]{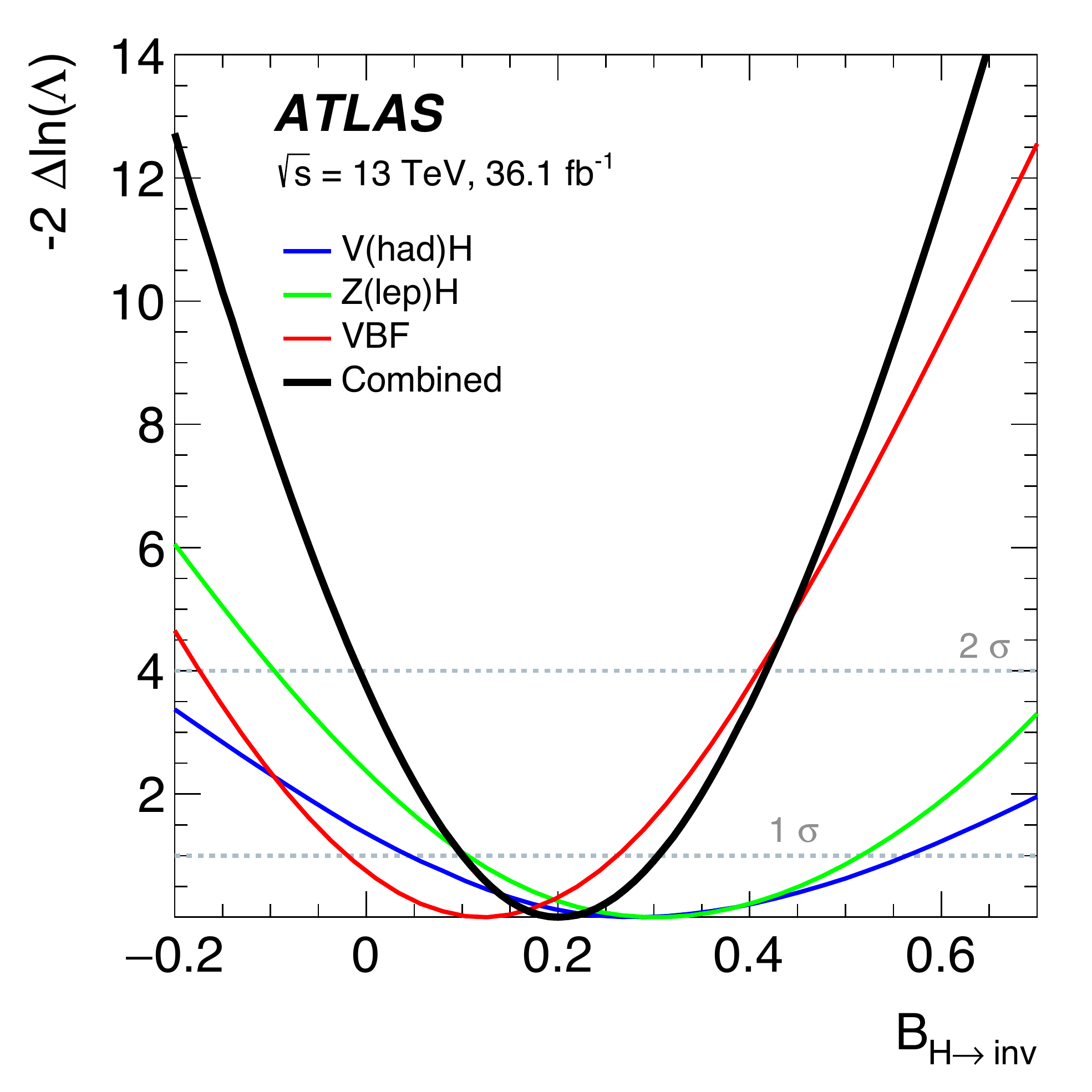}\qquad
\includegraphics[width=0.45\textwidth]{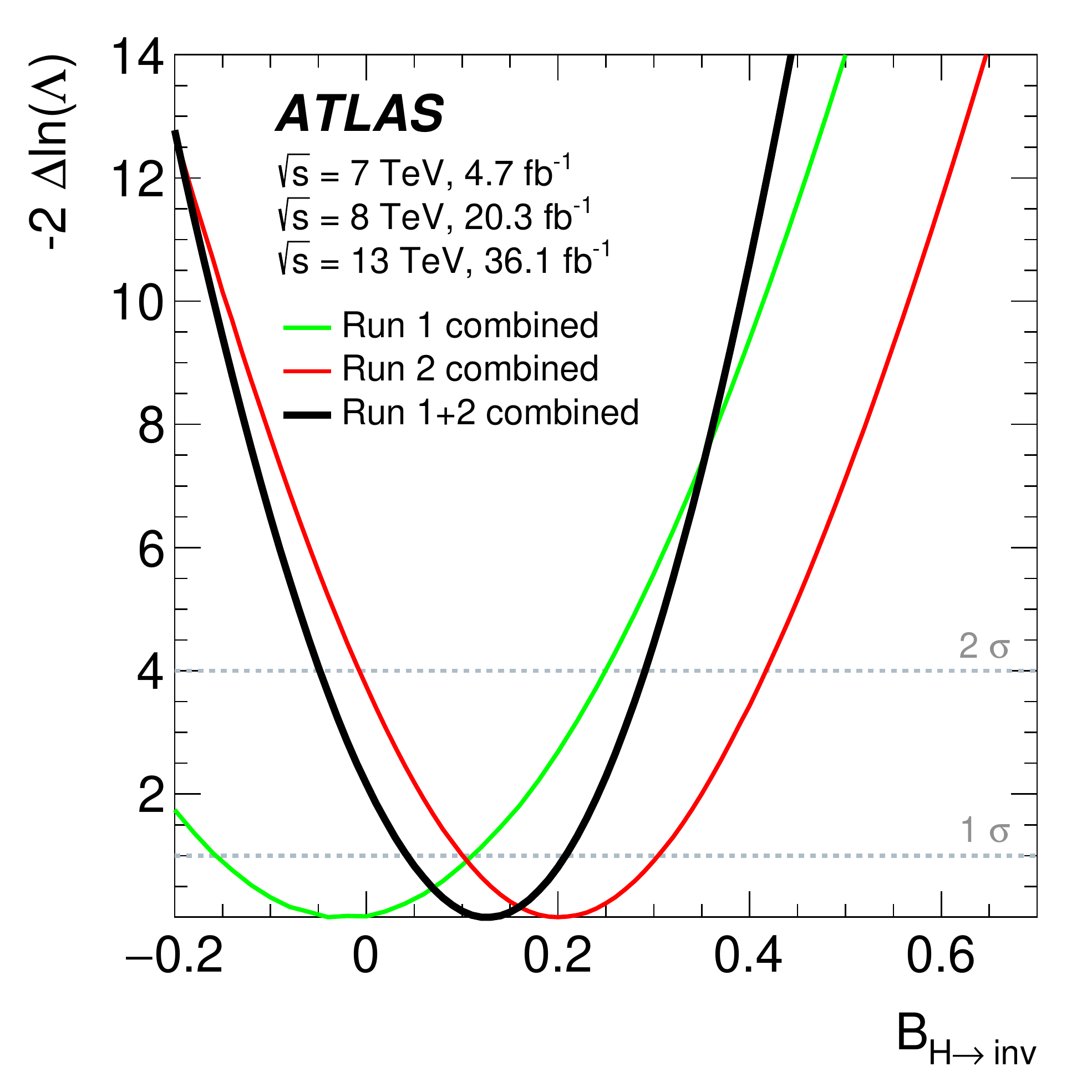}
\caption{
The observed negative logarithmic profile likelihood ratios $-2\,\Delta\!\ln(\Lambda)$ as a function of \brinv of the $\vh$, $\zh$, and VBF topologies using Run~2 data only and their statistical combination~(left). The $-2\,\Delta\!\ln(\Lambda)$ functions for the Run~2 combination together with the Run~1 combination and the total Run 1+2 combination~(right).
\label{fig:lhscan}
}
\end{figure}

Subsequently, the above Run 2 result is combined with the Run 1 searches for \hinv decays~\cite{bib:run1paper}.
Due to the differences between the detector layouts and data-taking conditions,  reconstruction algorithms and their calibrations, and  treatment of systematic uncertainties, the correlations between the runs  are not clearly identifiable. Hence, no correlations between Run 1 and 2 are assumed for most instrumental uncertainties.
The uncertainties related to the modeling of the calorimeter response dependence on jet flavor and pile-up are taken as either correlated or uncorrelated between the runs, and the choice which results in a weaker expected exclusion limit on \brinv is adopted.
The uncertainty on the JES of $b$-quark jets was estimated using MC simulations~\cite{PERF-2012-01,PERF-2016-04} and is therefore considered correlated.
For the signal modeling, the parton shower uncertainty in the \vh channel, the uncertainty from missing higher order corrections in the \zh analysis, and the uncertainty on the jet multiplicity in the VBF channel~\cite{Stewart:2011cf} are each taken as correlated between the runs since the estimated uncertainties stem from the same source.
For the same reason, the uncertainty from missing higher order corrections on the \met observable in the dominant background from diboson production in the \zh search is treated as correlated.
All other background modeling uncertainties are considered uncorrelated.
The impact of these correlation assumptions on the combined \brinv limit is found to be at most 0.005.
In addition, the impact on \brinv in scenarios ranging from full anti-correlation to full correlation is studied using the Best Linear Unbiased Estimator~(BLUE)~\cite{Lyons:1988rp} for the components of the JES uncertainty, the \vjets background, and  diboson production that are nominally not correlated due to different parameterisations in Run~1 and~2. The resulting absolute effect on the \brinv limit is at most 0.01.
 
The observed  $-2\,\Delta\!\ln(\Lambda)(\brinv;\theta)$ ratio of the combined Run~1+2 result is represented in Fig.~\ref{fig:lhscan}, alongside the individual Run~1 and Run~2 combinations. A best-fit value of $\brinv=0.13\pm0.08$ is obtained, corresponding to an observed~(expected) upper limit of $\brinv~<~\resultall$ at 95\%~CL.
The $p_\text{SM}$-value under the SM hypothesis is 10\%, and the compatibility between the Run~1 and Run~2 results is $1.5$ standard deviations. The final result, together with the results in the individual Run~2 analyses as well as the Run~2-only and the Run~1-only combinations, are summarized in Table~\ref{tab:resultsall}, and the upper limits on \brinv are graphically represented in Fig.~\ref{fig:combo_run1}. The results are consistent with a similar statistical combination in Ref.~\cite{Sirunyan:2018owy}.

\begin{table}
{
\caption{
Observed and expected upper limits  on \brinv at 95\% CL from direct searches for invisible decays of the 125~\GeV\ Higgs boson and statistical combinations. Also given are the observed $p$-values under the SM hypothesis.
\label{tab:resultsall}
}
\renewcommand{\arraystretch}{1.3}
\begin{center}
\resizebox{\textwidth}{!}{
\begin{tabular}{l|llllllll}
\hline
\hline
Analysis  & $\sqrt{s}$ & Int.\ luminosity & Observed & Expected  & $p_\text{SM}$-value &  Reference  \\
\hline
Run 2 VBF  & 13~TeV & 36.1~\ifb & $0.37$ & $0.28^{+0.11}_{-0.08}$ & 0.19 & \cite{bib:vbfpaper} \\
Run 2 $\zh$ & 13~TeV & 36.1~\ifb & $0.67$ & $0.39^{+0.17}_{-0.11}$ & 0.06 & \cite{bib:zhpaper} \\
Run 2 $\vh$ & 13~TeV & 36.1~\ifb & $0.83$ & $0.58^{+0.23}_{-0.16}$ &  0.12 & \cite{bib:vhpaper} \\
\hline
Run 2 Comb. & 13~TeV & 36.1~\ifb & 0.38 & $0.21^{+0.08}_{-0.06}$ & 0.03 & this Letter \\
Run 1 Comb. & $7,8$~TeV & 4.7, 20.3~\ifb & $0.25$ & $0.27^{+0.10}_{-0.08}$ & --- & \cite{bib:run1paper} \\
\hline
Run 1+2 Comb. & $7,8,13$~TeV & $4.7, 20.3, 36.1~\ifb$ & 0.26 & $0.17^{+0.07}_{-0.05}$ & 0.10 & this Letter \\
\hline
\hline
\end{tabular}
}
\end{center}
\renewcommand{\arraystretch}{1.0}
}
\end{table}

\begin{figure}
\centering
\includegraphics[width=0.65\textwidth]{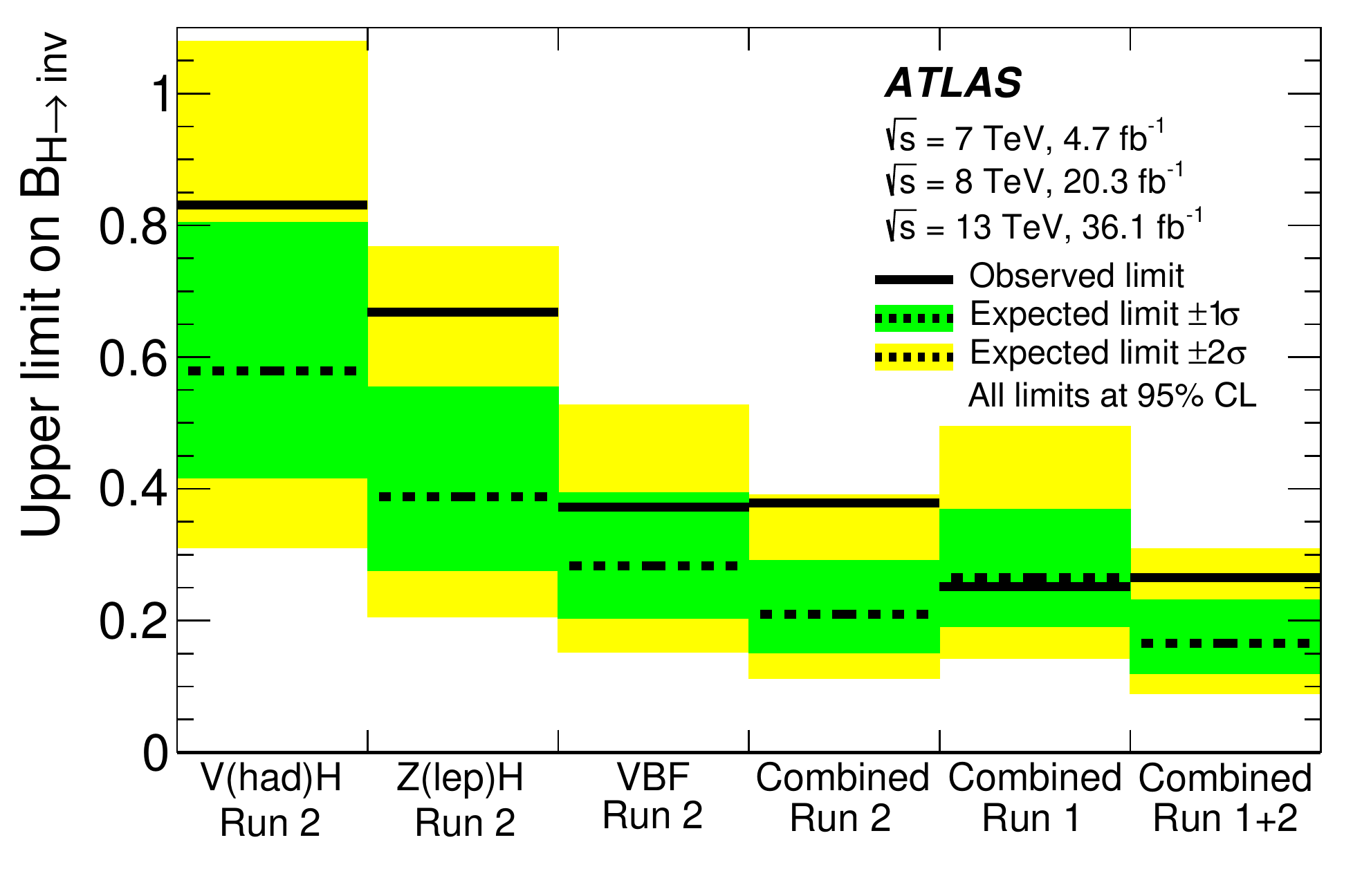}
\caption{
The observed and expected upper limits  on \brinv at 95\% CL from direct searches for invisible decays of the 125~\GeV\ Higgs boson and their statistical combinations in Run 1 and 2.
\label{fig:combo_run1}
}
\end{figure}

The constraint from the combined observed Run~1+2 exclusion limit of $\brinv < 0.24$
at 90\% CL is compared to the results from representative direct DM detection experiments~\cite{Petricca:2017zdp,Akerib:2016vxi,Cui:2017nnn,Aprile:2018dbl,Agnes:2018ves} in Fig.~\ref{fig:dd}.
This comparison is performed in the context of Higgs portal models~\cite{Patt:2006fw}.
The translation of the \hinv result into a weak interacting massive particle–nucleon
scattering cross section \sigmawn relies on an effective field theory approach~\cite{HIGG-2013-16}
under the assumption that invisible Higgs decays to a pair of WIMPs are kinematically possible and that
the WIMP is a scalar or a fermion~\cite{Eboli:2000ze,Fox:2011pm,deSimone:2014pda}, using the nuclear form factor $f_N=0.308\pm0.018$~\cite{Hoferichter:2017olk}.
The excluded \sigmawn values range down to $2 \times10^{-45}~$cm$^{2}$ in the scalar WIMP scenario. In the fermion WIMP case, the effective coupling is reduced by $m_H^2$~\cite{HIGG-2013-16}, excluding \sigmawn values down to $10^{-46}~$cm$^{2}$.
While the ATLAS exclusion limits extend to $m_\text{WIMP}<1~\GeV$, that region is subject to uncertainties in modelling of the nuclear recoil and is therefore not included in Figure~\ref{fig:dd}.

\begin{figure}
\centering
\includegraphics[width=0.7\textwidth]{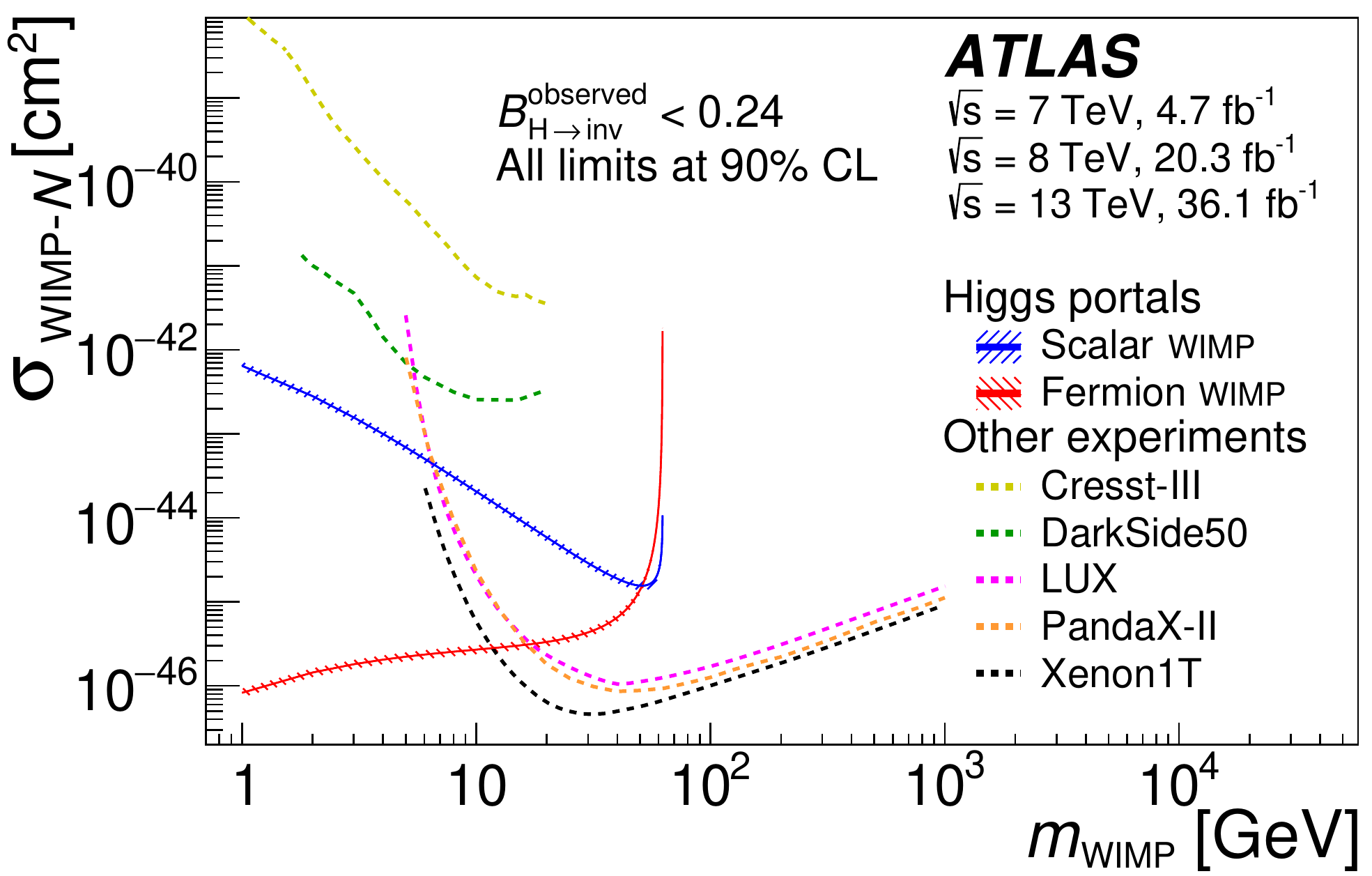}
\caption{
Comparison of the upper limits at 90\% CL from direct detection experiments~\cite{Petricca:2017zdp,Akerib:2016vxi,Cui:2017nnn,Aprile:2018dbl,Agnes:2018ves} on the spin-independent WIMP-nucleon scattering cross section to the observed exclusion limits from this analysis, assuming Higgs portal scenarios where the 125~GeV Higgs boson decays to a pair of DM particles~\cite{Patt:2006fw,HIGG-2013-16}. The regions above the limit contours are excluded in the range shown in the plot.
\label{fig:dd}
}
\end{figure}

In summary, direct searches for invisible Higgs boson decays using 36.1~\ifb\ of $pp$ collision data at $\sqrt{s} = 13~\TeV$ recorded in 2015 and 2016 in the VBF, $\zh$, and $\vh$ topologies are statistically combined assuming SM-like Higgs boson production. An upper limit on the invisible Higgs branching ratio of
$
\brinv{}<\result
$
is observed~(expected) at 95\% CL. A statistical combination of this result with the combination of direct \hinv searches using up to 4.7~\ifb\ of $pp$ collision data at $\sqrt{s} = 7~\TeV$ and up to 20.3~\ifb\ at 8~TeV collected in Run~1 of the LHC yields an observed~(expected) upper limit of
$
\brinv{}<\resultall
$
at 95\% CL.
The combined Run~1+2 result is translated into upper limits on the WIMP–nucleon scattering cross section for Higgs portal models.
The derived limits range down to $2 \times 10^{-45}~$cm$^{2}$ in the scalar and $10^{-46}~$cm$^{2}$ in the fermion WIMP scenarios, highlighting the complementarity of DM searches at the LHC and direct detection experiments.

% The next lines are included from the .//acknowledgements/Acknowledgements.tex input file
 
We thank CERN for the very successful operation of the LHC, as well as the
support staff from our institutions without whom ATLAS could not be
operated efficiently.
 
We acknowledge the support of ANPCyT, Argentina; YerPhI, Armenia; ARC, Australia; BMWFW and FWF, Austria; ANAS, Azerbaijan; SSTC, Belarus; CNPq and FAPESP, Brazil; NSERC, NRC and CFI, Canada; CERN; CONICYT, Chile; CAS, MOST and NSFC, China; COLCIENCIAS, Colombia; MSMT CR, MPO CR and VSC CR, Czech Republic; DNRF and DNSRC, Denmark; IN2P3-CNRS, CEA-DRF/IRFU, France; SRNSFG, Georgia; BMBF, HGF, and MPG, Germany; GSRT, Greece; RGC, Hong Kong SAR, China; ISF and Benoziyo Center, Israel; INFN, Italy; MEXT and JSPS, Japan; CNRST, Morocco; NWO, Netherlands; RCN, Norway; MNiSW and NCN, Poland; FCT, Portugal; MNE/IFA, Romania; MES of Russia and NRC KI, Russian Federation; JINR; MESTD, Serbia; MSSR, Slovakia; ARRS and MIZ\v{S}, Slovenia; DST/NRF, South Africa; MINECO, Spain; SRC and Wallenberg Foundation, Sweden; SERI, SNSF and Cantons of Bern and Geneva, Switzerland; MOST, Taiwan; TAEK, Turkey; STFC, United Kingdom; DOE and NSF, United States of America. In addition, individual groups and members have received support from BCKDF, CANARIE, CRC and Compute Canada, Canada; COST, ERC, ERDF, Horizon 2020, and Marie Sk{\l}odowska-Curie Actions, European Union; Investissements d' Avenir Labex and Idex, ANR, France; DFG and AvH Foundation, Germany; Herakleitos, Thales and Aristeia programmes co-financed by EU-ESF and the Greek NSRF, Greece; BSF-NSF and GIF, Israel; CERCA Programme Generalitat de Catalunya, Spain; The Royal Society and Leverhulme Trust, United Kingdom.
 
The crucial computing support from all WLCG partners is acknowledged gratefully, in particular from CERN, the ATLAS Tier-1 facilities at TRIUMF (Canada), NDGF (Denmark, Norway, Sweden), CC-IN2P3 (France), KIT/GridKA (Germany), INFN-CNAF (Italy), NL-T1 (Netherlands), PIC (Spain), ASGC (Taiwan), RAL (UK) and BNL (USA), the Tier-2 facilities worldwide and large non-WLCG resource providers. Major contributors of computing resources are listed in Ref.~\cite{ATL-GEN-PUB-2016-002}.
 
% End of text imported from the .//acknowledgements/Acknowledgements.tex input file

\clearpage

\printbibliography

@article{Grazzini:2015hta,
      author         = "Grazzini, Massimiliano and Kallweit, Stefan and Rathlev,
                        Dirk",
      title          = "{ZZ production at the LHC: fiducial cross sections and
                        distributions in NNLO QCD}",
      journal        = "Phys. Lett. B",
      volume         = "750",
      year           = "2015",
      pages          = "407-410",
      doi            = "10.1016/j.physletb.2015.09.055",
      eprint         = "1507.06257",
      archivePrefix  = "arXiv",
      primaryClass   = "hep-ph",
      SLACcitation   = "%%CITATION = ARXIV:1507.06257;%%"
}

@article{Caola:2015psa,
      author         = "Caola, Fabrizio and Melnikov, Kirill and Rontsch, Raoul
                        and Tancredi, Lorenzo",
      title          = "{QCD corrections to ZZ production in gluon fusion at the
                        LHC}",
      journal        = "Phys. Rev. D",
      volume         = "92",
      year           = "2015",
      number         = "9",
      pages          = "094028",
      doi            = "10.1103/PhysRevD.92.094028",
      eprint         = "1509.06734",
      archivePrefix  = "arXiv",
      primaryClass   = "hep-ph",
      reportNumber   = "TTP15-035, CERN-PH-TH-2015-226, FERMILAB-PUB-15-398-T",
      SLACcitation   = "%%CITATION = ARXIV:1509.06734;%%"
}

@article{Goldberg:1983nd,
      author         = "Goldberg, H.",
      title          = "{Constraint on the Photino Mass from Cosmology}",
      journal        = "Phys. Rev. Lett.",
      volume         = "50",
      year           = "1983",
      pages          = "1419",
      doi            = "10.1103/PhysRevLett.50.1419",
      note           = "Erratum: Phys. Rev. Lett. \textbf{103} (2009) 099905",
      reportNumber   = "NUB-2592",
      SLACcitation   = "%%CITATION = PRLTA,50,1419;%%"
}

@article{Ellis:1983ew,
      author         = "Ellis, John R. and Hagelin, J. S. and Nanopoulos, Dimitri
                        V. and Olive, Keith A. and Srednicki, M.",
      title          = "{Supersymmetric Relics from the Big Bang}",
      booktitle      = "{IN *BATAVIA 1984, PROCEEDINGS, INNER SPACE/OUTER SPACE*,
                        458-459., In *Srednicki, M.A. (ed.): Particle physics and
                        cosmology* 223-246}",
      journal        = "Nucl. Phys. B",
      volume         = "238",
      year           = "1984",
      pages          = "453",
      doi            = "10.1016/0550-3213(84)90461-9",
      reportNumber   = "SLAC-PUB-3171",
      SLACcitation   = "%%CITATION = NUPHA,B238,453;%%"
}

@article{Akrami:2018vks,
      author         = "{Planck Collaboration}",
      title          = "{Planck 2018 results. I. Overview and the cosmological
                        legacy of Planck}",
      collaborations  = "Planck",
      year           = "2018",
      eprint         = "1807.06205",
      archivePrefix  = "arXiv",
      primaryClass   = "astro-ph.CO",
      SLACcitation   = "%%CITATION = ARXIV:1807.06205;%%"
}

@article{Hinshaw:2012aka,
      author         = "{WMAP Collaboration}",
      pauthor         = "Hinshaw, G. and others",
      title          = "{Nine-Year Wilkinson Microwave Anisotropy Probe (WMAP)
                        Observations: Cosmological Parameter Results}",
      collaborations  = "WMAP",
      journal        = "Astrophys. J. Suppl.",
      volume         = "208",
      year           = "2013",
      pages          = "19",
      doi            = "10.1088/0067-0049/208/2/19",
      eprint         = "1212.5226",
      archivePrefix  = "arXiv",
      primaryClass   = "astro-ph.CO",
      SLACcitation   = "%%CITATION = ARXIV:1212.5226;%%"
}

@article{Trible:dmu,
author = {Trimble, Virginia},
title = {Existence and Nature of Dark Matter in the Universe},
      journal        = "Ann. Rev. Astron. Astrophys.",
volume = {25},
number = {1},
pages = {425-472},
year = {1987},
doi = {10.1146/annurev.aa.25.090187.002233},
pURL = {https://doi.org/10.1146/annurev.aa.25.090187.002233},
peprint = {https://doi.org/10.1146/annurev.aa.25.090187.002233},
}

@article{Feng:2010gw,
      author         = "Feng, Jonathan L.",
      title          = "{Dark Matter Candidates from Particle Physics and Methods
                        of Detection}",
      journal        = "Ann. Rev. Astron. Astrophys.",
      volume         = "48",
      year           = "2010",
      pages          = "495-545",
      doi            = "10.1146/annurev-astro-082708-101659",
      eprint         = "1003.0904",
      archivePrefix  = "arXiv",
      primaryClass   = "astro-ph.CO",
      reportNumber   = "UCI-TR-2009-13",
      SLACcitation   = "%%CITATION = ARXIV:1003.0904;%%"
}

@article{Boveia:2018yeb,
      author         = "Boveia, Antonio and Doglioni, Caterina",
      title          = "{Dark Matter Searches at Colliders}",
      journal        = "Ann. Rev. Nucl. Part. Sci.",
      volume         = "68",
      year           = "2018",
      pages          = "429-459",
      doi            = "10.1146/annurev-nucl-101917-021008",
      eprint         = "1810.12238",
      archivePrefix  = "arXiv",
      primaryClass   = "hep-ex",
      SLACcitation   = "%%CITATION = ARXIV:1810.12238;%%"
}

@article{Sirunyan:2018owy,
      author         = "{CMS Collaboration}",
      title          = "{Search for invisible decays of a Higgs boson produced
                        through vector boson fusion in proton-proton collisions at
                        $\sqrt{s} =$ 13 TeV}",
      collaboration  = "CMS",
      journal        = "accepted by Phys. Lett. B",
      year           = "2018",
      eprint         = "1809.05937",
      doi            = "10.1016/j.physletb.2019.04.025",
      archivePrefix  = "arXiv",
      primaryClass   = "hep-ex",
      reportNumber   = "CMS-HIG-17-023, CERN-EP-2018-139",
      SLACcitation   = "%%CITATION = ARXIV:1809.05937;%%"
}

@article{Sirunyan:2017qfc,
      author         = "{CMS Collaboration}",
      title          = "{Search for new physics in events with a leptonically
                        decaying Z boson and a large transverse momentum imbalance
                        in proton–proton collisions at $\sqrt{s} $ = 13 $\,\text
                        {TeV}$}",
      collaboration  = "CMS",
      journal        = "Eur. Phys. J. C",
      volume         = "78",
      year           = "2018",
      number         = "4",
      pages          = "291",
      doi            = "10.1140/epjc/s10052-018-5740-1",
      eprint         = "1711.00431",
      archivePrefix  = "arXiv",
      primaryClass   = "hep-ex",
      reportNumber   = "CMS-EXO-16-052, CERN-EP-2017-259,
                        FERMILAB-PUB-17-534-CMS",
      SLACcitation   = "%%CITATION = ARXIV:1711.00431;%%"
}

@article{Sirunyan:2017jix,
      author         = "{CMS Collaboration}",
      title          = "{Search for new physics in final states with an energetic
                        jet or a hadronically decaying $W$ or $Z$ boson and
                        transverse momentum imbalance at $\sqrt{s}=13\text{
                        }\text{ }\mathrm{TeV}$}",
      collaboration  = "CMS",
      journal        = "Phys. Rev. D",
      volume         = "97",
      year           = "2018",
      number         = "9",
      pages          = "092005",
      doi            = "10.1103/PhysRevD.97.092005",
      eprint         = "1712.02345",
      archivePrefix  = "arXiv",
      primaryClass   = "hep-ex",
      reportNumber   = "CMS-EXO-16-048, CERN-EP-2017-294",
      SLACcitation   = "%%CITATION = ARXIV:1712.02345;%%"
}

@article{Khachatryan:2016whc,
      author         = "{CMS Collaboration}",
      title          = "{Searches for invisible decays of the Higgs boson in pp
                        collisions at $\sqrt{s}$ = 7, 8, and 13 TeV}",
      collaboration  = "CMS",
      journal        = "JHEP",
      volume         = "02",
      year           = "2017",
      pages          = "135",
      doi            = "10.1007/JHEP02(2017)135",
      eprint         = "1610.09218",
      archivePrefix  = "arXiv",
      primaryClass   = "hep-ex",
      reportNumber   = "CMS-HIG-16-016, CERN-EP-2016-240",
      SLACcitation   = "%%CITATION = ARXIV:1610.09218;%%"
}

@article{Chatrchyan:2014tja,
      author         = "{CMS Collaboration}",
      title          = "{Search for invisible decays of Higgs bosons in the
                        vector boson fusion and associated ZH production modes}",
      collaboration  = "CMS",
      journal        = "Eur. Phys. J. C",
      volume         = "74",
      year           = "2014",
      pages          = "2980",
      doi            = "10.1140/epjc/s10052-014-2980-6",
      eprint         = "1404.1344",
      archivePrefix  = "arXiv",
      primaryClass   = "hep-ex",
      reportNumber   = "CMS-HIG-13-030, CERN-PH-EP-2014-051",
      SLACcitation   = "%%CITATION = ARXIV:1404.1344;%%"
}

@article{Lyons:1988rp,
      author         = "Lyons, Louis and Gibaut, Duncan and Clifford, Peter",
      title          = "{How to Combine Correlated Estimates of a Single Physical
                        Quantity}",
      journal        = "Nucl. Instrum. Meth.",
      volume         = "A270",
      year           = "1988",
      pages          = "110",
      doi            = "10.1016/0168-9002(88)90018-6",
      reportNumber   = "PRINT-88-0048 (OXFORD), OXFORD-NP-5-88",
      SLACcitation   = "%%CITATION = NUIMA,A270,110;%%"
}

@article{Stewart:2011cf,
      author         = "Stewart, Iain W. and Tackmann, Frank J.",
      title          = "{Theory Uncertainties for Higgs and Other Searches Using
                        Jet Bins}",
      journal        = "Phys. Rev. D",
      volume         = "85",
      year           = "2012",
      pages          = "034011",
      doi            = "10.1103/PhysRevD.85.034011",
      eprint         = "1107.2117",
      archivePrefix  = "arXiv",
      primaryClass   = "hep-ph",
      SLACcitation   = "%%CITATION = ARXIV:1107.2117;%%"
}

@inproceedings{Petricca:2017zdp,
      author         = "Petricca, F. and others",
      title          = "{First results on low-mass dark matter from the
                        CRESST-III experiment}",
      booktitle      = "{15th International Conference on Topics in Astroparticle
                        and Underground Physics (TAUP 2017) Sudbury, Ontario,
                        Canada, July 24-28, 2017}",
      collaboration  = "CRESST",
      year           = "2017",
      eprint         = "1711.07692",
      archivePrefix  = "arXiv",
      primaryClass   = "astro-ph.CO",
      SLACcitation   = "%%CITATION = ARXIV:1711.07692;%%"
}

@article{Akerib:2016vxi,
      author         = "Akerib, D. S. and others",
      title          = "{Results from a search for dark matter in the complete
                        LUX exposure}",
      collaboration  = "LUX",
      journal        = "Phys. Rev. Lett.",
      volume         = "118",
      year           = "2017",
      number         = "2",
      pages          = "021303",
      doi            = "10.1103/PhysRevLett.118.021303",
      eprint         = "1608.07648",
      archivePrefix  = "arXiv",
      primaryClass   = "astro-ph.CO",
      SLACcitation   = "%%CITATION = ARXIV:1608.07648;%%"
}

@article{Cui:2017nnn,
      author         = "Cui, Xiangyi and others",
      title          = "{Dark Matter Results From 54-Ton-Day Exposure of
                        PandaX-II Experiment}",
      collaboration  = "PandaX-II",
      journal        = "Phys. Rev. Lett.",
      volume         = "119",
      year           = "2017",
      number         = "18",
      pages          = "181302",
      doi            = "10.1103/PhysRevLett.119.181302",
      eprint         = "1708.06917",
      archivePrefix  = "arXiv",
      primaryClass   = "astro-ph.CO",
      SLACcitation   = "%%CITATION = ARXIV:1708.06917;%%"
}

@article{Aprile:2018dbl,
      author         = "Aprile, E. and others",
      title          = "{Dark Matter Search Results from a One Ton-Year Exposure
                        of XENON1T}",
      collaboration  = "XENON",
      journal        = "Phys. Rev. Lett.",
      volume         = "121",
      year           = "2018",
      number         = "11",
      pages          = "111302",
      doi            = "10.1103/PhysRevLett.121.111302",
      eprint         = "1805.12562",
      archivePrefix  = "arXiv",
      primaryClass   = "astro-ph.CO",
      SLACcitation   = "%%CITATION = ARXIV:1805.12562;%%"
}

@article{Agnes:2018ves,
      author         = "Agnes, P. and others",
      title          = "{Low-Mass Dark Matter Search with the DarkSide-50
                        Experiment}",
      collaboration  = "DarkSide",
      journal        = "Phys. Rev. Lett.",
      volume         = "121",
      year           = "2018",
      number         = "8",
      pages          = "081307",
      doi            = "10.1103/PhysRevLett.121.081307",
      eprint         = "1802.06994",
      archivePrefix  = "arXiv",
      primaryClass   = "astro-ph.HE",
      reportNumber   = "FERMILAB-PUB-18-048-AE-E-PPD",
      SLACcitation   = "%%CITATION = ARXIV:1802.06994;%%"
}

@article{Hoferichter:2017olk,
      author         = "Hoferichter, Martin and Klos, Philipp and Men\'e{}ndez,
                        Javier and Schwenk, Achim",
      title          = "{Improved Limits for Higgs-Portal Dark Matter from LHC Searches}",
      journal        = "Phys. Rev. Lett.",
      volume         = "119",
      year           = "2017",
      number         = "18",
      pages          = "181803",
      doi            = "10.1103/PhysRevLett.119.181803",
      eprint         = "1708.02245",
      archivePrefix  = "arXiv",
      primaryClass   = "hep-ph",
      reportNumber   = "INT-PUB-17-031",
      SLACcitation   = "%%CITATION = ARXIV:1708.02245;%%"
}

@article{bib:run1paper,
      author         = "{ATLAS Collaboration}",
      title          = "{Constraints on new phenomena via Higgs boson couplings
                        and invisible decays with the ATLAS detector}",
      collaboration  = "ATLAS",
      journal        = "JHEP",
      volume         = "11",
      year           = "2015",
      pages          = "206",
      doi            = "10.1007/JHEP11(2015)206",
      eprint         = "1509.00672",
      archivePrefix  = "arXiv",
      primaryClass   = "hep-ex",
      reportNumber   = "CERN-PH-EP-2015-191",
      SLACcitation   = "%%CITATION = ARXIV:1509.00672;%%"
}

@article{bib:vbfpaper,
      author         = "{ATLAS Collaboration}",
      title          = "{Search for invisible Higgs boson decays in vector boson
                        fusion at $\sqrt{s} = 13$ TeV with the ATLAS detector}",
      pcollaboration  = "ATLAS",
      year           = "2018",
      eprint         = "1809.06682",
      archivePrefix  = "arXiv",
      primaryClass   = "hep-ex",
      reportNumber   = "CERN-EP-2018-184",
      SLACcitation   = "%%CITATION = ARXIV:1809.06682;%%"
}

@article{bib:zhpaper,
      author         = "{ATLAS Collaboration}",
      title          = "{Search for an invisibly decaying Higgs boson or dark
                        matter candidates produced in association with a $Z$ boson
                        in $pp$ collisions at $\sqrt{s} =$ 13 TeV with the ATLAS
                        detector}",
      journal        = "Phys. Lett. B",
      volume         = "776",
      year           = "2018",
      pages          = "318-337",
      doi            = "10.1016/j.physletb.2017.11.049",
      eprint         = "1708.09624",
      archivePrefix  = "arXiv",
      primaryClass   = "hep-ex",
      reportNumber   = "CERN-EP-2017-166",
      SLACcitation   = "%%CITATION = ARXIV:1708.09624;%%"
}

@article{bib:vhpaper,
      author         = "{ATLAS Collaboration}",
      pauthor         = "Aaboud, M. and others",
      title          = "{Search for dark matter in events with a hadronically
                        decaying vector boson and missing transverse momentum in
                        $pp$ collisions at $\sqrt{s} = 13$ TeV with the ATLAS
                        detector}",
      collaboration  = "ATLAS",
      journal        = "JHEP",
      volume         = "10",
      year           = "2018",
      pages          = "180",
      doi            = "10.1007/JHEP10(2018)180",
      eprint         = "1807.11471",
      archivePrefix  = "arXiv",
      primaryClass   = "hep-ex",
      reportNumber   = "CERN-EP-2018-083",
      SLACcitation   = "%%CITATION = ARXIV:1807.11471;%%"
}

@article{Jungman:1995df,
      author         = "Jungman, Gerard and Kamionkowski, Marc and Griest, Kim",
      title          = "{Supersymmetric dark matter}",
      journal        = "Phys. Rept.",
      volume         = "267",
      year           = "1996",
      pages          = "195",
      doi            = "10.1016/0370-1573(95)00058-5",
      eprint         = "hep-ph/9506380",
      archivePrefix  = "arXiv",
      reportNumber   = "SU-4240-605, UCSD-PTH-95-02, IASSNS-HEP-95-14, CU-TP-677",
      SLACcitation   = "%%CITATION = HEP-PH/9506380;%%"
}

@article{Scherrer:1985zt,
      author         = "Scherrer, Robert J. and Turner, Michael S.",
      title          = "{On the relic, cosmic abundance of stable, weakly interacting massive particles}",
      journal        = "Phys. Rev. D",
      volume         = "33",
      year           = "1986",
      pages          = "1585",
      doi            = "10.1103/PhysRevD.33.1585",
      note           = "[Erratum: Phys. Rev. D \textbf{34} (1986) 3263]",
      reportNumber   = "FERMILAB-PUB-85-163-A, EFI-85-76-CHICAGO",
      SLACcitation   = "%%CITATION = PHRVA,D33,1585;%%"
}

@article{ATLAS-CONF-2016-035,
      author         = "{ATLAS Collaboration}",
      title         = "{Jet mass reconstruction with the ATLAS Detector in early
                       Run 2 data}",
      institution   = "CERN",
      collaboration = "ATLAS Collaboration",
      address       = "Geneva",
      journal       = "ATLAS-CONF-2016-035",
      month         = "Jul",
      year          = "2016",
      reportNumber  = "ATLAS-CONF-2016-035",
      url           = "https://cds.cern.ch/record/2200211",
}

@article{Capeans:2010jnh,
      author  = "{ATLAS Collaboration}",
      pauthor         = "Capeans, M. and Darbo, G. and Einsweiler, K. and Elsing,
                        M. and Flick, T. and Garcia-Sciveres, M. and Gemme, C. and
                        Pernegger, H. and Rohne, O. and Vuillermet, R.",
      title          = "{ATLAS Insertable B-Layer Technical Design Report}",
      year           = "2010",
      journal = "ATLAS-TDR-19",
      purl          = "http://cds.cern.ch/record/1291633",
      note = {\textsc{url: }\href{http://cds.cern.ch/record/1291633}{http://cds.cern.ch/record/1291633}, ATLAS Insertable B-Layer Technical Design Report Addendum, ATLAS-TDR-19-ADD-1 (2012), \textsc{url: }\href{http://cds.cern.ch/record/1451888}{http://cds.cern.ch/record/1451888}},
      prelated = {ATLAS:2012rar},
      prelatedstring={Corrected in},
      reportNumber   = "CERN-LHCC-2010-013, ATLAS-TDR-19",
      SLACcitation   = "%%CITATION = CERN-LHCC-2010-013;%%",
}

@article{Abbott:2018ikt,
      author         = "Abbott, B. and others",
      title          = "{Production and Integration of the ATLAS Insertable
                        B-Layer}",
      collaboration  = "ATLAS IBL",
      journal        = "JINST",
      volume         = "13",
      year           = "2018",
      number         = "05",
      pages          = "T05008",
      doi            = "10.1088/1748-0221/13/05/T05008",
      eprint         = "1803.00844",
      archivePrefix  = "arXiv",
      primaryClass   = "physics.ins-det",
      SLACcitation   = "%%CITATION = ARXIV:1803.00844;%%"
}

@article{Abercrombie:2015wmb,
      author         = "Abercrombie, Daniel and others",
      title          = "{Dark Matter Benchmark Models for Early LHC Run-2
                        Searches: Report of the ATLAS/CMS Dark Matter Forum}",
      year           = "2015",
      eprint         = "1507.00966",
      archivePrefix  = "arXiv",
      primaryClass   = "hep-ex",
      reportNumber   = "FERMILAB-PUB-15-282-CD",
      SLACcitation   = "%%CITATION = ARXIV:1507.00966;%%"
}

@article{Bertone:2004pz,
      author         = "Bertone, Gianfranco and Hooper, Dan and Silk, Joseph",
      title          = "{Particle dark matter: Evidence, candidates and
                        constraints}",
      journal        = "Phys. Rept.",
      volume         = "405",
      year           = "2005",
      pages          = "279",
      doi            = "10.1016/j.physrep.2004.08.031",
      eprint         = "hep-ph/0404175",
      archivePrefix  = "arXiv",
      reportNumber   = "FERMILAB-PUB-04-047-A",
      SLACcitation   = "%%CITATION = HEP-PH/0404175;%%"
}

@article{Steigman:1984ac,
      author         = "Steigman, Gary and Turner, Michael S.",
      title          = "Cosmological constraints on the properties of weakly
                        interacting massive particles",
      journal        = "Nucl. Phys. B",
      volume         = "253",
      year           = "1985",
      pages          = "375",
      doi            = "10.1016/0550-3213(85)90537-1",
      reportNumber   = "FERMILAB-PUB-84-110-A, BA-84-33",
      SLACcitation   = "%%CITATION = NUPHA,B253,375;%%"
}

@article{Patt:2006fw,
      author         = "Patt, Brian and Wilczek, Frank",
      title          = "{Higgs-field portal into hidden sectors}",
      year           = "2006",
      eprint         = "hep-ph/0605188",
      archivePrefix  = "arXiv",
      primaryClass   = "hep-ph",
      reportNumber   = "MIT-CTP-3745",
      SLACcitation   = "%%CITATION = HEP-PH/0605188;%%"
}

@article{Fox:2011pm,
      author         = "Fox, Patrick J. and Harnik, Roni and Kopp, Joachim and
                        Tsai, Yuhsin",
      title          = "{Missing Energy Signatures of Dark Matter at the LHC}",
      journal        = "Phys. Rev. D",
      volume         = "85",
      year           = "2012",
      pages          = "056011",
      doi            = "10.1103/PhysRevD.85.056011",
      eprint         = "1109.4398",
      archivePrefix  = "arXiv",
      primaryClass   = "hep-ph",
      reportNumber   = "FERMILAB-PUB-11-478-T",
      SLACcitation   = "%%CITATION = ARXIV:1109.4398;%%"
}

@article{deSimone:2014pda,
      author         = "De Simone, Andrea and Giudice, Gian Francesco and
                        Strumia, Alessandro",
      title          = "{Benchmarks for Dark Matter Searches at the LHC}",
      journal        = "JHEP",
      volume         = "06",
      year           = "2014",
      pages          = "081",
      doi            = "10.1007/JHEP06(2014)081",
      eprint         = "1402.6287",
      archivePrefix  = "arXiv",
      primaryClass   = "hep-ph",
      reportNumber   = "CERN-PH-TH-2014-008, SISSA-01-2014-FISI",
      SLACcitation   = "%%CITATION = ARXIV:1402.6287;%%"
}

@article{Aad:2012tfa,
      author         = "{ATLAS Collaboration}",
      title          = "{Observation of a new particle in the search for the
                        Standard Model Higgs boson with the ATLAS detector at the
                        LHC}",
      collaboration  = "ATLAS",
      journal        = "Phys. Lett. B",
      volume         = "716",
      year           = "2012",
      pages          = "1",
      doi            = "10.1016/j.physletb.2012.08.020",
      eprint         = "1207.7214",
      archivePrefix  = "arXiv",
      primaryClass   = "hep-ex",
      reportNumber   = "CERN-PH-EP-2012-218",
      SLACcitation   = "%%CITATION = ARXIV:1207.7214;%%"
}

@article{Chatrchyan:2012ufa,
      author         = "{CMS Collaboration}",
      title          = "{Observation of a new boson at a mass of 125 GeV with the
                        CMS experiment at the LHC}",
      collaboration  = "CMS Collaboration",
      journal        = "Phys. Lett. B",
      volume         = "716",
      pages          = "30",
      doi            = "10.1016/j.physletb.2012.08.021",
      year           = "2012",
      eprint         = "1207.7235",
      archivePrefix  = "arXiv",
      primaryClass   = "hep-ex",
      reportNumber   = "CMS-HIG-12-028, CERN-PH-EP-2012-220",
      SLACcitation   = "%%CITATION = ARXIV:1207.7235;%%",
}

@article{Read:2002hq,
     author         = "Read, Alexander L.",
      title          = "{Presentation of search results: the CL$_{s}$ technique}",
      journal        = "J. Phys. G",
      volume         = "28",
      pages          = "2693",
      doi            = "10.1088/0954-3899/28/10/313",
      year           = "2002",
      SLACcitation   = "%%CITATION = INSPIRE-599622;%%",
}

@article{Aad:2008zzm,
      author         = "{ATLAS Collaboration}",
      title          = "{The ATLAS Experiment at the CERN Large Hadron Collider}",
      collaboration  = "ATLAS",
      journal        = "JINST",
      volume         = "3",
      year           = "2008",
      pages          = "S08003",
      doi            = "10.1088/1748-0221/3/08/S08003",
      SLACcitation   = "%%CITATION = JINST,3,S08003;%%"
}

@article{Cacciari:2008gp,
      author         = "Cacciari, Matteo and Salam, Gavin P. and Soyez, Gregory",
      title          = "{The anti-$k_t$ jet clustering algorithm}",
      journal        = "JHEP",
      volume         = "04",
      year           = "2008",
      pages          = "063",
      doi            = "10.1088/1126-6708/2008/04/063",
      eprint         = "0802.1189",
      archivePrefix  = "arXiv",
      primaryClass   = "hep-ph",
      reportNumber   = "LPTHE-07-03",
      SLACcitation   = "%%CITATION = ARXIV:0802.1189;%%"
}

@article{Krohn2010,
	author         = "Krohn, David and Thaler, Jesse and Wang, Lian-Tao",
      title          = "{Jet trimming}",
      journal        = "JHEP",
      volume         = "02",
      year           = "2010",
      pages          = "084",
      doi            = "10.1007/JHEP02(2010)084",
      eprint         = "0912.1342",
      archivePrefix  = "arXiv",
      primaryClass   = "hep-ph",
      SLACcitation   = "%%CITATION = ARXIV:0912.1342;%%"}

@article{Cowan:2010js,
      author         = "Cowan, Glen and others",
      title          = "{Asymptotic formulae for likelihood-based tests of new
                        physics}",
      journal        = "Eur. Phys. J. C",
      volume         = "71",
      pages          = "1554",
      doi            = "10.1140/epjc/s10052-011-1554-0",
      note           = "[Erratum: Eur. Phys. J. C  \textbf{73} (2013) 2501]",
      year           = "2011",
      eprint         = "1007.1727",
      archivePrefix  = "arXiv",
      primaryClass   = "physics.data-an",
      SLACcitation   = "%%CITATION = ARXIV:1007.1727;%%",
}

@article{Verkerke:2003ir,
      author         = "Verkerke, Wouter and Kirkby, David P.",
      title          = "{The RooFit toolkit for data modeling}",
      booktitle      = "{Statistical Problems in Particle Physics, Astrophysics
                        and Cosmology (PHYSTAT 05): Proceedings, Oxford, UK,
                        September 12-15, 2005}",
      journal        = "eConf",
      volume         = "C0303241",
      year           = "2003",
      pages          = "MOLT007",
      pnote           = "[,186(2003)]",
      eprint         = "physics/0306116",
      archivePrefix  = "arXiv",
      primaryClass   = "physics",
      reportNumber   = "CHEP-2003-MOLT007",
      SLACcitation   = "%%CITATION = PHYSICS/0306116;%%"
}

@Booklet{ATL-GEN-PUB-2016-002,
    author         = "{ATLAS Collaboration}",
    title          = "{ATLAS Computing Acknowledgements}",
    howpublished   = "{ATL-GEN-PUB-2016-002}",
    url            = "https://cds.cern.ch/record/2202407",
}
 
\clearpage
% ATLAS Collaboration author list
% Reference date of HIGG-2018-54 is 2018-11-12
% Author list last updated on date 10-APR-19
% Data extracted on 10-Apr-2019 for paper reference HIGG-2018-54
% at 10:04am
 
\begin{flushleft}
{\Large The ATLAS Collaboration}

\bigskip

M.~Aaboud$^\textrm{\scriptsize 35d}$,    
G.~Aad$^\textrm{\scriptsize 101}$,    
B.~Abbott$^\textrm{\scriptsize 128}$,    
D.C.~Abbott$^\textrm{\scriptsize 102}$,    
O.~Abdinov$^\textrm{\scriptsize 13,*}$,    
A.~Abed~Abud$^\textrm{\scriptsize 70a,70b}$,    
D.K.~Abhayasinghe$^\textrm{\scriptsize 93}$,    
S.H.~Abidi$^\textrm{\scriptsize 167}$,    
O.S.~AbouZeid$^\textrm{\scriptsize 40}$,    
N.L.~Abraham$^\textrm{\scriptsize 156}$
% ...
\bigskip
\\

% ...
$^{13}$Institute of Physics, Azerbaijan Academy of Sciences, Baku; Azerbaijan.\\
% ...
$^{35}$$^{(a)}$Facult\'e des Sciences Ain Chock, R\'eseau Universitaire de Physique des Hautes Energies - Universit\'e Hassan II, Casablanca;$^{(b)}$Facult\'{e} des Sciences, Universit\'{e} Ibn-Tofail, K\'{e}nitra;$^{(c)}$Facult\'e des Sciences Semlalia, Universit\'e Cadi Ayyad, LPHEA-Marrakech;$^{(d)}$Facult\'e des Sciences, Universit\'e Mohamed Premier and LPTPM, Oujda;$^{(e)}$Facult\'e des sciences, Universit\'e Mohammed V, Rabat; Morocco.\\
% ...
$^{40}$Niels Bohr Institute, University of Copenhagen, Copenhagen; Denmark.\\
% ...
$^{70}$$^{(a)}$INFN Sezione di Pavia;$^{(b)}$Dipartimento di Fisica, Universit\`a di Pavia, Pavia; Italy.\\
% ...
$^{93}$Department of Physics, Royal Holloway University of London, Egham; United Kingdom.\\
% ...
$^{101}$CPPM, Aix-Marseille Universit\'e, CNRS/IN2P3, Marseille; France.\\
$^{102}$Department of Physics, University of Massachusetts, Amherst MA; United States of America.\\
% ...
$^{128}$Homer L. Dodge Department of Physics and Astronomy, University of Oklahoma, Norman OK; United States of America.\\
% ...
$^{156}$Department of Physics and Astronomy, University of Sussex, Brighton; United Kingdom.\\
% ...
$^{167}$Department of Physics, University of Toronto, Toronto ON; Canada.\\
% ...
$^{a}$ Also at Centre for High Performance Computing, CSIR Campus, Rosebank, Cape Town; South Africa.\\
$^{b}$ Also at CERN, Geneva; Switzerland.\\
$^{c}$ Also at CPPM, Aix-Marseille Universit\'e, CNRS/IN2P3, Marseille; France.\\
$^{d}$ Also at D\'epartement de Physique Nucl\'eaire et Corpusculaire, Universit\'e de Gen\`eve, Gen\`eve; Switzerland.\\
$^{e}$ Also at Departament de Fisica de la Universitat Autonoma de Barcelona, Barcelona; Spain.\\
% ...
$^{*}$ Deceased

\end{flushleft}

% Created with Glance <Atlas.Glance@cern.ch>

\end{document}